\documentclass{PoS}

\usepackage[english]{babel}
\usepackage[utf8]{inputenc}
\usepackage[T1]{fontenc}

\title{Lattice Determination of the Anomalous Magnetic Moment of the Muon}

\ShortTitle{Lattice Determination of the Anomalous Magnetic Moment of the Muon}

\author{Michele Della Morte, \speaker{Benjamin Jäger}, Hartmut Wittig\\
Institut für Kernphysik and Helmholtz Institut Mainz,\\
Johannes Gutenberg-Universität, D-55099 Mainz, Germany\\
       E-mail: \email{morte@kph.uni-mainz.de},
       \email{jaeger@kph.uni-mainz.de}, \email{wittig@kph.uni-mainz.de}}

\author{Andreas Jüttner\\
        CERN, Physics Department, TH Unit, CH-1211 Geneva 23, Switzerland\\
       E-mail: \email{Andreas.Juttner@cern.ch}}

\abstract{\vspace{-11cm} \phantom{a} \hfill MKPH-T-11-21 \newline \phantom{a} \hfill HIM-2011-11 \newline \phantom{a} \hfill CERN-PH-TH-2011-280 \newline  \vspace{9cm} \newline 
We compute the leading hadronic contribution to the anomalous magnetic moment of the muon $a_\mu^\mathrm{HLO}$ using two dynamical
flavours of non-perturbatively $\mathcal{O}(a)$ improved Wilson fermions. By applying
partially twisted boundary conditions we are able to improve the momentum resolution of the vacuum
polarisation, an important ingredient for the determination of the leading
hadronic contribution. We check systematic uncertainties by studying several
ensembles, which allows us to discuss finite size effects and lattice artefacts.
The chiral behavior of $a_\mu^\mathrm{HLO}$ turns out to be non-trivial, especially for
small pion masses.}

\FullConference{XXIX International Symposium on Lattice Field Theory \\
                July 10 – 16 2011\\
                Squaw Valley, Lake Tahoe, California}

\begin{document}

\section{Introduction}

The anomalous magnetic moment of the muon is defined as half the difference
between the gyromagnetic factor of the muon $g_\mu$ and 2: $a_\mu
= (g_\mu-2) / 2$. This quantity is commonly used to test the Standard Model of
particle physics, since it can be measured and computed to a very high
precision. Currently a discrepancy of $3.2$
standard deviations between experiment and theory is observed~[1], which might be a hint
for physics beyond the Standard Model. The hadronic contributions, especially
the hadronic vacuum polarisation $a_\mu^\mathrm{HLO}$, dominate the theoretical
uncertainties. Currently the Standard Model prediction of $a_\mu^\mathrm{HLO}$
is determined using the optical theorem to relate the cross-section
$e^+e^-\,\rightarrow\,\mathit{hadrons}$ data to the vacuum polarisation. A
calculation from first principles is clearly desirable. Here we report on our current effort  towards a precise determination of the leading hadronic contribution to the anomalous magnetic moment of the muon using Lattice {Q}{C}{D}.

\section{Lattice Setup}

\begin{table}[h!]
\begin{center}

\begin{tabular}{ccccccc}
    	\hline
    	$\beta$ & $a$ $[\mathrm{fm}]$ & lattice & $L$ $[\mathrm{fm}]$ & 
    	$m_\pi$ $[\mathrm{MeV}]$ & $m_\pi L$ & Labels\\
    	\hline
    	$5.20$ & $0.079$ & $64 \times 32^3$ & $2.5$ & $471$ - $317$ &  $6.0$ -
    	$4.0$ & A3 - A5 \\
    	\hline
    	$5.30$ & $0.063$ & $64 \times 32^3$ & $2.0$ & $644$ - $447$ &   $7.9$ -
    	$4.7$ & E3 - E5 \\
    	$5.30$ & $0.063$ & $96 \times 48^3$ & $3.0$ & $323$, $277$ &  $5.0$, $4.2$
    	& F6, F7 \\
    	\hline
    	$5.50$ & $0.050$ & $96 \times 48^3$ & $2.4$ & $541$, $431$  &  $6.5$,
    	$5.2$ & N4, N5 \\
    	\hline
\end{tabular}
\caption{Summary of simulations parameters. Measurements are performed on
configurations separated by $8$ units of molecular dynamics time at least. The
scale and pion masses are still preliminary and taken from~[10,11].}
\label{tab}
\end{center}
\end{table}
We use two dynamical flavours of
non-perturbatively $\mathcal{O}(a)$ improved Wilson fermions~[2] and include a partially quenched strange quark.
Similar studies have been performed in the quenched approximation~[3,4] and in
the theory with two~[5,6] and three dynamical flavours~[7,8]. Our measurements
are performed on a subset of the gauge configurations generated as part of the CLS
project~[9]. The simulations parameters are listed in Table \ref{tab}. On the
lattice the vacuum polarisation tensor is defined by a Fourier-transformation
of a current-current correlator
\begin{equation}
\Pi_{\mu \nu}(q) = a^4 \sum_{x} e^{i q (x + a\hat{\mu}/2 - a
\hat{\nu}/2)} \left< J^\mathrm{c}_\mu(x) J^\mathrm{c}_\nu(0) \right> ,
\label{PImunu} \end{equation} where we use the conserved point-split current
$J^\mathrm{c}_\mu(x)$~[4] given in the case of Wilson fermions by 
\begin{equation}
J^\mathrm{c}_\mu (x) = \frac{1}{2} \bigg(  \bar{q}(x+a\hat{\mu})
(1+\gamma_\mu) U_\mu^+(x) q(x) - \bar{q}(x) (1- \gamma_\mu) U_\mu(x)
q(x+a\hat{\mu}) \bigg).
\end{equation} Preforming the Wick contractions in Equation \ref{PImunu}
produces connected as well as disconnected contributions. Disconnected
diagrams are computationally expensive and neglected in this study. Nevertheless
two-flavour chiral perturbation theory of NLO allows us to estimate the
disconnected contributions to be $-10\%$ of the connected one~[12].
 
Current conservation implies that the vacuum polarisation tensor
$\Pi_{\mu \nu}(q)$  can be related to the vacuum polarisation $\Pi(q^2)$ in the
following way:
\begin{equation}
\Pi_{\mu \nu}(q) =\left( \delta_{\mu\nu}
q^2 - q_\mu q_\nu \right) \Pi(q^2) .
\end{equation} 
For space-like momenta, the relation between the vacuum polarisation $\Pi(q^2)$
and the lowest order hadronic contribution  to the anomalous
magnetic moment of the muon $a_\mu^\mathrm{HLO}$ has been derived in [3,4,13]
\begin{equation}
a_\mu = \left(\frac{\alpha}{\pi}\right) ^2 \int_0^\infty dq^2\,f(q^2)
\hat{\Pi}(q^2), \label{amu-int}
\end{equation}
where the kernel of this integral is given by
\begin{equation}
f(q^2) = \frac{m_\mu^2 q^2 Z^3 (1- q^2 Z)}{1+m_\mu^2 q^2 Z^2} , \qquad
Z=\frac{q^2- \sqrt{q^4 - 4 m_\mu^2 q^2}}{2 m_\mu^2 q^2}
\end{equation}
and $\hat{\Pi}(q^2) = 4 \pi \left(\Pi(q^2) - \Pi(0)  \right)$. 
We have implemented partially twisted boundary conditions~[14]
\begin{equation}
q(x+L\,\hat{k}) = e^{i \theta_k} q(x) ,
\end{equation}
which allow us to access any value of the momentum $\frac{2\pi}{L} \vec{n} -
\frac{\vec{\theta}}{L}$, where $\vec{n}$ is a vector of integers. In this way we
are  able to improve the sampling with data points, in particular the
kinematical region $q^2 < \left( \frac{2\pi}{L} \right)^2$ where the kernel of
the integral in Equation~\ref{amu-int} is peaked. Partially twisted boundary
conditions can be applied to the connected diagram by reinterpreting the correlator as 
flavour non-diagonal (see~[15]).


\section{Determination of $a_\mu^\mathrm{HLO}$}

In order to determine the leading order hadronic contribution to $a_\mu$ via the
integral in Equation~\ref{amu-int}, we need a continuous description of the
vacuum polarisation $\Pi(q^2)$. Therefore we perform correlated least square
fits to our simulation data. Perturbation theory including
$\mathcal{O}(\alpha_s)$ terms~[16] is incorporated into the fitting procedure to
constrain the fit at large momentum by demanding that the fit-function and
perturbation theory are matched at some high momentum. To ensure a smooth function, we also imply a matching of the first derivative at the same point. This procedure reduces the number of
free fit coefficients by two. To evaluate the perturbative formula we use the
non-perturbative, two-flavour Lambda-parameter $\Lambda_{\overline{MS}}$
parameter from~[17] and the non-perturbative renormalization factors
in~[18,19,20]. We study systematic effects introduced by the choice of the
fit procedure, by varying the fit ansatz and the matching point of perturbation theory. We chose 4 different fit-ansätze and checked for systematic differences:
\begin{enumerate}
  \item[a)] a model independent Pad\'e with degree $2$ over degree $3$ 
  \begin{equation}
  	\Pi(q^2)= \frac{a (q^2+b^2)(q^2+c^2)}{(q^2+d^2) (q^2+e^2) (q^2+f^2)} ,
  \end{equation}
  \item[b)] vector dominance model including a single vector
  \begin{equation}
  	\Pi(q^2)= a + \frac{b}{(q^2+c^2)} ,
  \end{equation}
    \item[c)] vector dominance model with two vectors and one mass fixed to
    $m_V$ as proposed in~[7,8]
  \begin{equation}
  	\Pi(q^2)= a + \frac{b}{(q^2+c^2)} + \frac{d}{(q^2+m_V^2)}, 
  \end{equation}
    \item[d)] vector dominance model with two free vector masses
  \begin{equation}
  	\Pi(q^2)= a + \frac{b}{(q^2+c^2)} + \frac{d}{(q^2+e^2)}.
  \end{equation}
\end{enumerate}
\begin{figure}[h!]
\begin{minipage}{0.58\linewidth}
	\centering
	\includegraphics[width=\linewidth]{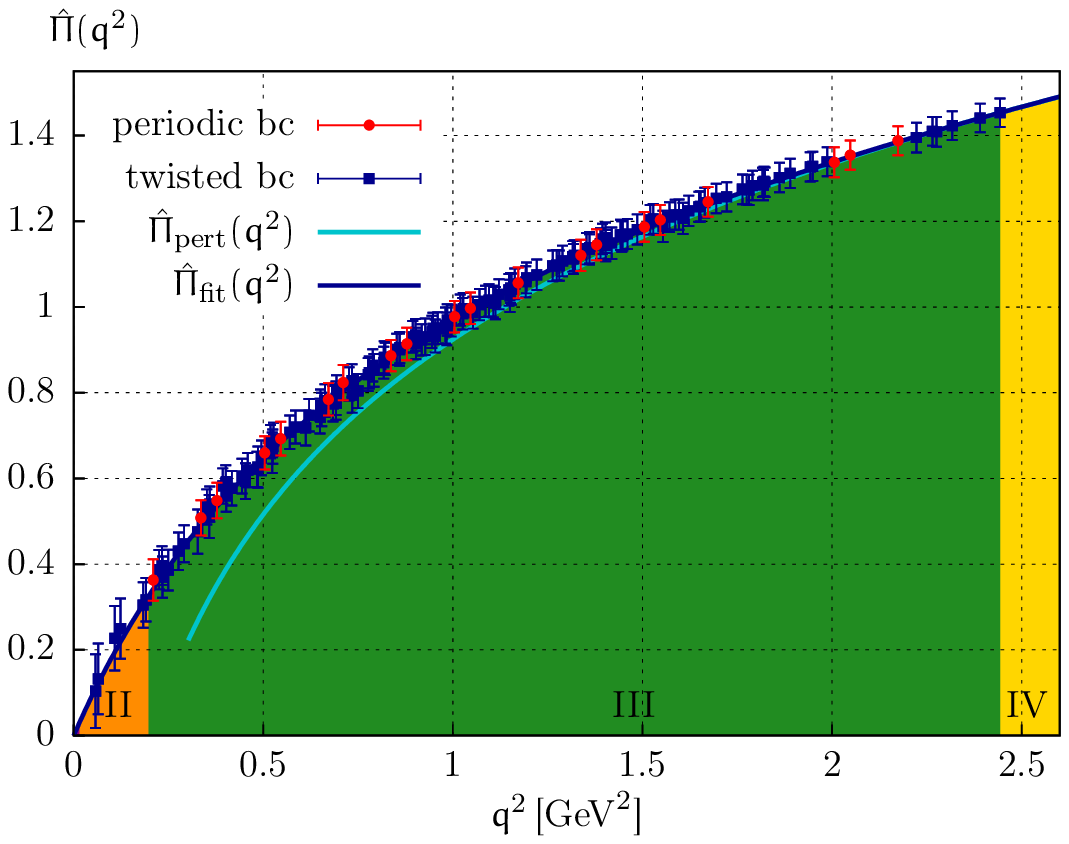}
\end{minipage}
\hfill
\begin{minipage}{0.38\linewidth}
	\centering
	\includegraphics[width=\linewidth]{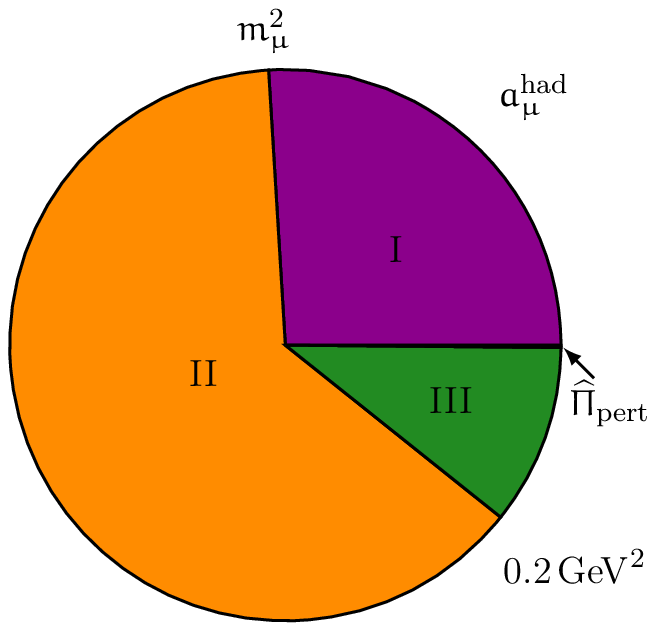}
\end{minipage}
\caption{{\bf Left:} The subtracted vacuum polarisation $\hat{\Pi}(q^2)$
computed on the F6 ensemble ($\beta=~5.3;$ $L=~3.0\,\mathrm{fm};$
$m_\pi=~323\,\mathrm{MeV}$) using twisted and periodic boundary conditions. The
blue solid line shows the fit to the double vector ansatz c) matched to
perturbation theory (light blue line). {\bf Right:} The different contributions
to $a_\mu^\mathrm{HLO}$ from the dispersion integral broken down to several momentum ranges, shown in both figures indicated by different colours.}
	\label{Bild1}
\end{figure}

The fits, except the vector dominance model fit, turn out to give consistent
values within the statistical uncertainties, when the matching point of
perturbation theory is chosen larger than $\approx 1\,\mathrm{GeV}$. In Figure~\ref{Bild1} we show the
subtracted vacuum polarisation together with the perturbation theory and fit
c) (i.e. double vector). The effect of twisting is illustrated by showing
periodic and twisted data in the same plot, which demonstrate a clear improvement on the momentum resolution of $\hat{\Pi}(q^2)$. The remaining integration of
Equation~\ref{amu-int} is performed numerically. The right panel of
Figure~\ref{Bild1} shows the individual contributions to $a_\mu^\mathrm{HLO}$
separated into different momentum regions. Part I shows the area for $0$ to $m_\mu^2$, in which
no data points occur. Nevertheless this small momentum range is constrained by
the condition that $\hat{\Pi}(q^2) \rightarrow 0$ for $q\rightarrow 0$
and the smooth behaviour of the fit curve. The second momentum range (II)
displays the region in which twisted data points begin to contribute. In the third region (III) periodic and
twisted data points give a perfect description of the momentum behaviour. The
final section (IV) shows the contribution from perturbation theory, which turns
out to be negligible. The overall statistical error, estimated via a bootstrap
procedure, is dominated by regions I and II and ranges from $2\%$ to $7\%$ for
the different ensembles.

\section{Results}

\begin{figure}[h!]
\begin{minipage}{0.49\linewidth}
	\centering
	\includegraphics[width=\linewidth]{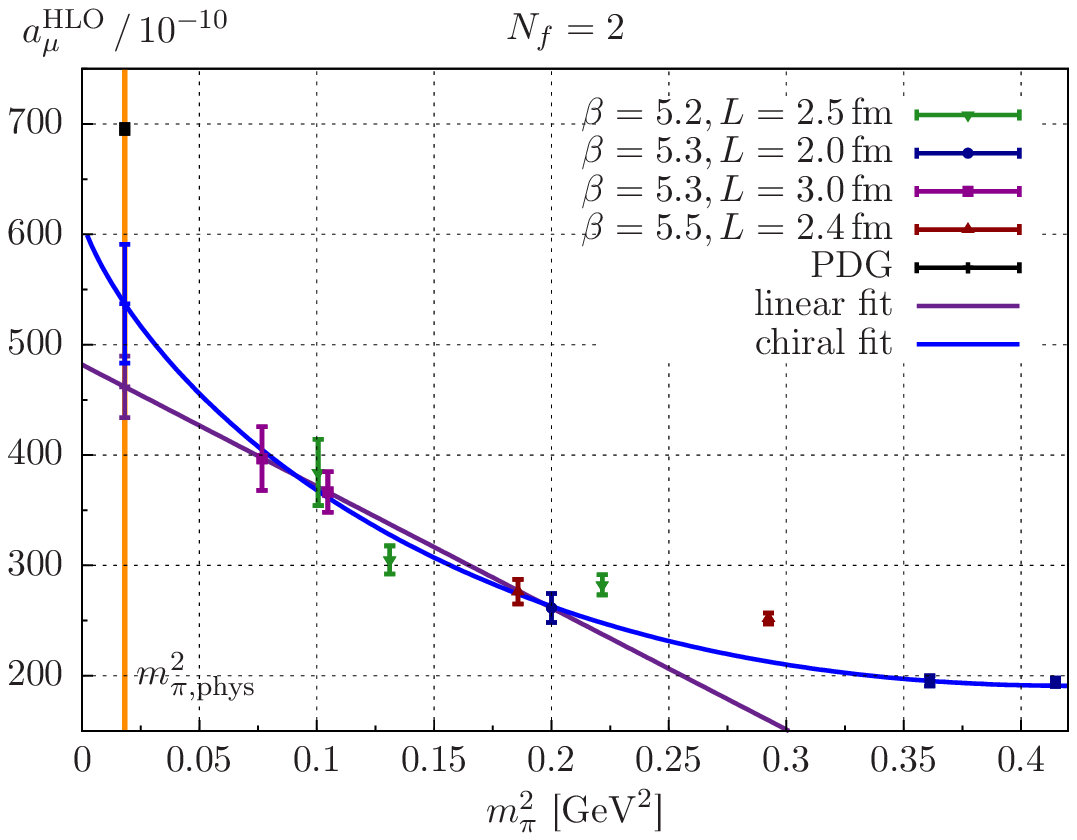}
\end{minipage}
\hfill
\begin{minipage}{0.49\linewidth}
	\centering
	\includegraphics[width=\linewidth]{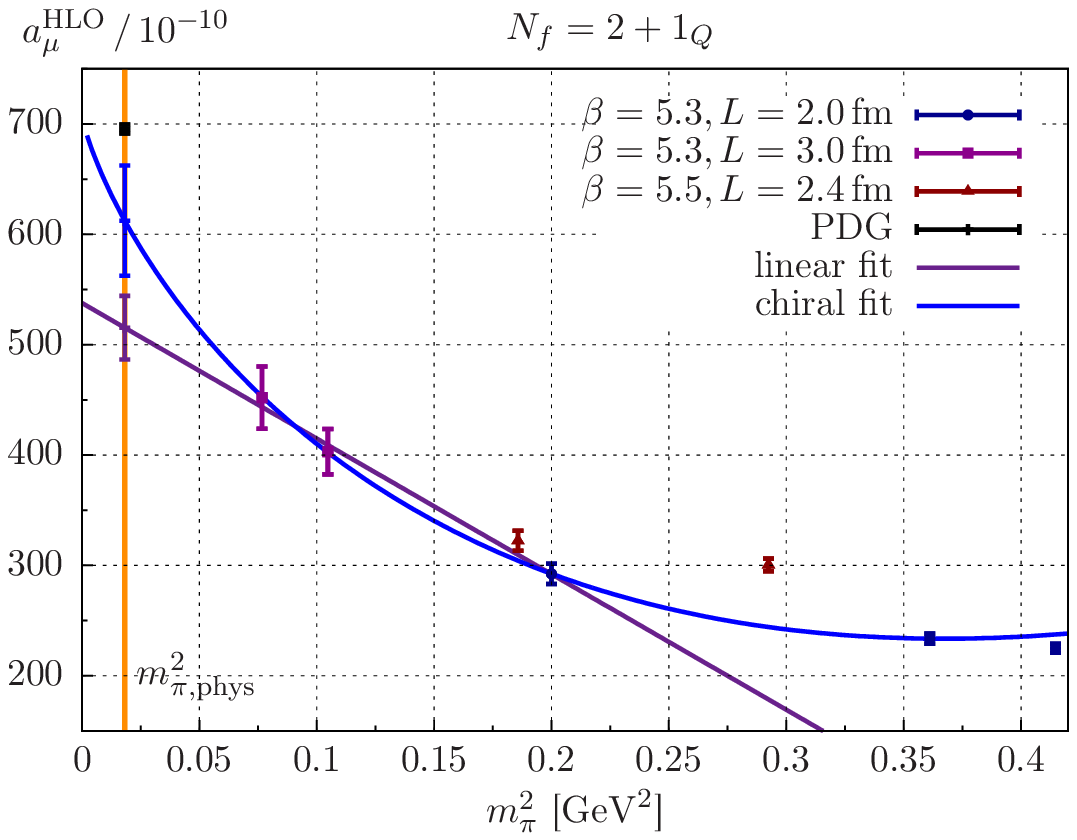}
\end{minipage}
\caption{{\bf Left:} The simulation results for the hadronic contribution
to $a_\mu$ computed using two flavours, shown as function of $m_\pi^2$. The
chiral extrapolation (blue curve) is performed using an ansatz motivated by
chiral perturbation theory. {\bf Right:} Corresponding results for
$a_\mu^\mathrm{HLO}$ including a partially quenched strange quark.}
\label{Bild2}
\end{figure}

In Figure~\ref{Bild2} we show the result for $a_\mu^\mathrm{HLO}$ computed for
the ensembles listed in Table~\ref{tab} as a function of $m_\pi^2$. We observe a
clear curvature and steep rise for small pion masses. To obtain
$a_\mu^\mathrm{HLO}$ at the physical point we need to perform a chiral
extrapolation. In addition to a linear behaviour in $m_\pi^2$ we also include a
logarithmic term, which is motivated by chiral perturbation theory
  \begin{equation}
  	a_\mu^\mathrm{HLO}(m_\pi^2)= a_\mu^\mathrm{HLO} + B\, m_\pi^2 + C\, m_\pi^2\, \log ( m_\pi^2) . 
  \end{equation}
We restrict the fit to the four most chiral $\beta=5.3$ data only to avoid
mixing of cutoff effects and the chiral extrapolation. 

We find that this extrapolation describes the whole set of data
points quite well even those which not included in the fit. If we use this
extrapolation, we obtain for $a_\mu^\mathrm{HLO}$ in the two-flavour case
\begin{equation}
a_\mu^\mathrm{HLO} (N_f=2) =  537.1\,
(53.8)_\mathrm{stat}\, (37.6)_\mathrm{chiral} \cdot 10^{-10} .
\end{equation}
For the case of an additional quenched strange quark we end up with 
\begin{equation}
a_\mu^\mathrm{HLO} (N_f=2+1_Q) = 612.4\,
(49.9)_\mathrm{stat}\, (48.5)_\mathrm{chiral} \cdot 10^{-10}.
\end{equation}

We repeat the analysis using a linear extrapolation for the $3$ most chiral
$\beta=5.3$ data points to estimate the uncertainties from the chiral
extrapolation by using half the difference of the central values. 

\begin{figure}[ht]
\begin{minipage}{0.49\linewidth}
	\centering
	\includegraphics[width=\linewidth]{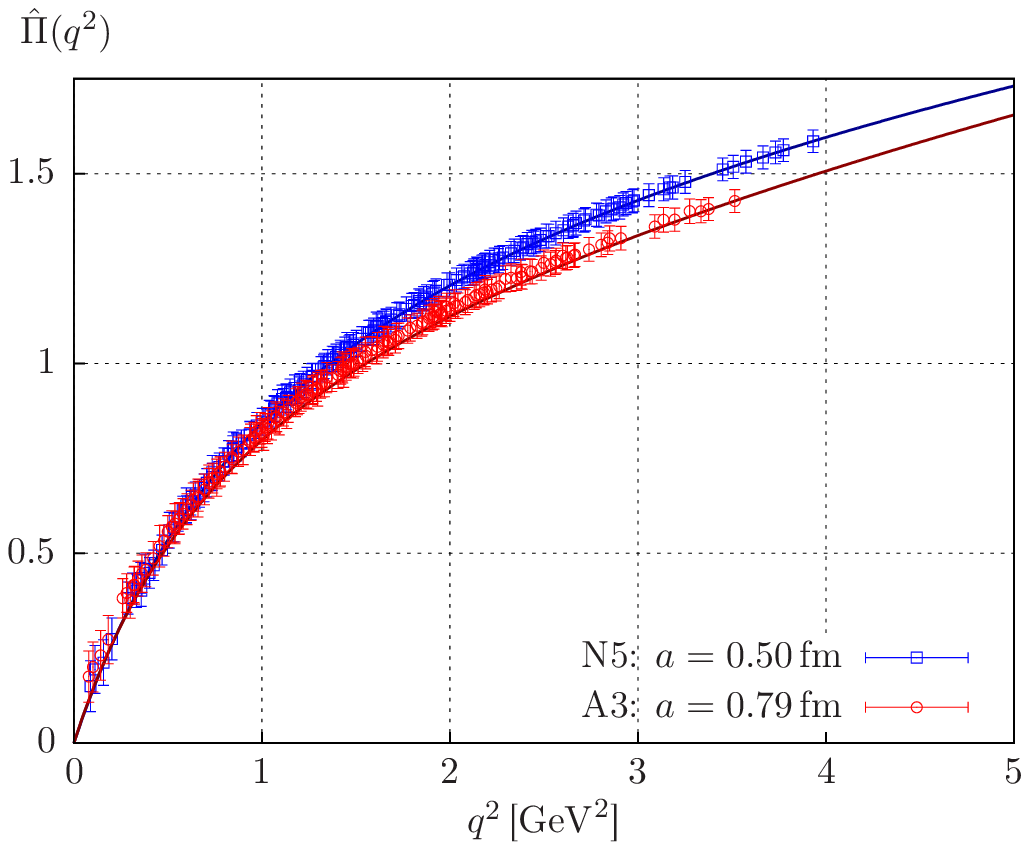}
\end{minipage}
\hfill
\begin{minipage}{0.49\linewidth}
	\centering
	\includegraphics[width=\linewidth]{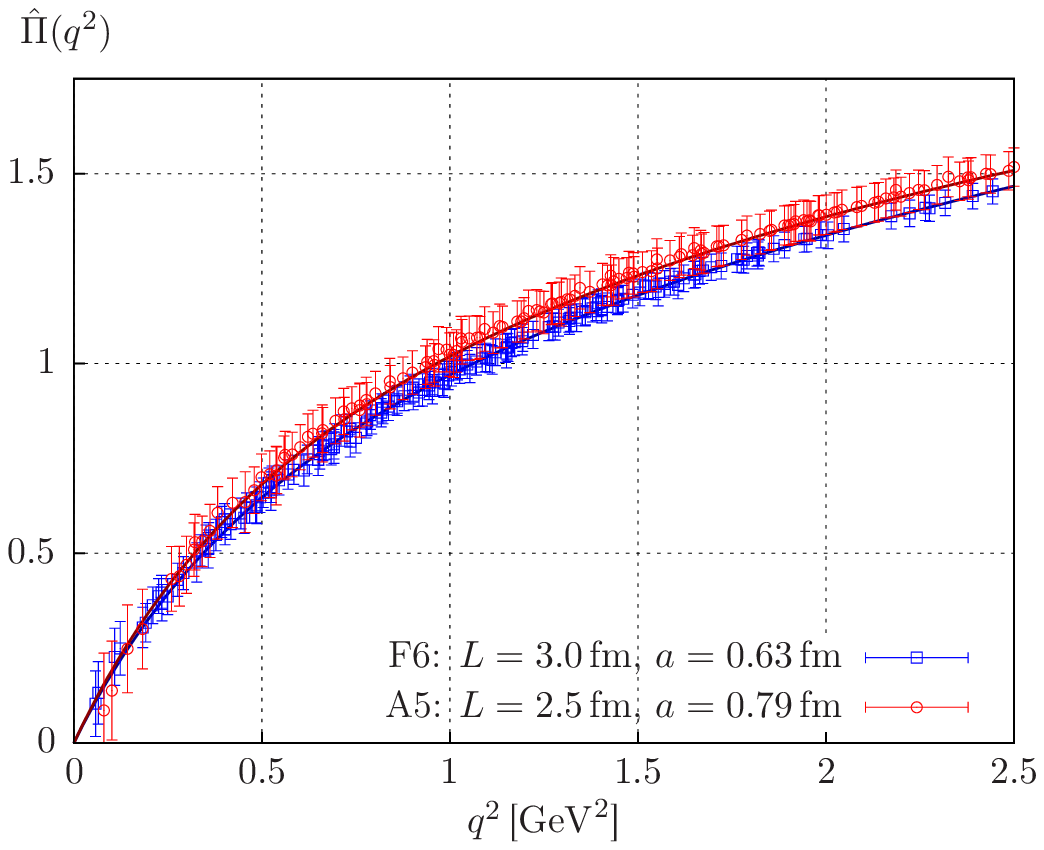}
\end{minipage}
\caption{{\bf Left:} A study of cutoff effects comparing the subtracted vacuum
polarisation for two ensemble with roughly the same pion mass $m_\pi\approx
450\,\mathrm{MeV}$ and volume $L=2.5\,\mathrm{fm}$. {\bf Right:} A comparison of
$\hat{\Pi}(q^2)$ for two ensemble with the same pion mass $m_\pi=320\,\mathrm{MeV}$.}
\label{Bild3}
\end{figure}

Since finite size effects and cutoff effects rather depend on $q^2$, it is more
instructive to study them using the subtracted vacuum polarisation
$\hat{\Pi}(q^2)$ than $a_\mu^\mathrm{HLO}$. The left panel of
Figure~\ref{Bild3} shows the quantity $\hat{\Pi}(q^2)$ for two
different ensembles which have roughly the same pion mass and volume. This allows us to look for cutoff effects, which
turn out to be below $3\%$ in the momentum range $q^2 < 1\,\mathrm{GeV}^2$. The
right panel of Figure~\ref{Bild3} offers an insight to finite size effects and cutoff effects, showing two ensemble which
differ in volume and lattice spacing. It turns out that these two effects are
around $5\%$ and within our statistical precision. 

\section{Conclusions and Outlook}

The determination of $a_\mu^\mathrm{HLO}$ using Lattice {Q}{C}{D} is feasible,
but still requires further improvements to make an impact on phenomenology.
Partially twisted boundary conditions extend the accessible momentum range for the vacuum polarisation and thereby reduce the statistical and systematic uncertainties. At present the individual data points for $a_\mu^\mathrm{HLO}$ have statistical precision of $2\%$ to $7\%$.
Summing up all individual uncertainties in quadrature we end up with an overall
uncertainty of $ \approx 12\%$ for the extrapolated value at the physical point.
Our study of residual systematics indicate that finite size effects and cutoff
effects change $a_\mu^\mathrm{HLO}$ slightly. The value will in any case
decrease after the inclusion of the disconnected diagrams. Further details of
our study will be published soon~[21]. In the future we will study ensembles
with smaller pion masses to improve the extrapolation to the physical point and reduce the systematics, such as finite size effects and cutoff effects. Once we
are confident that we can control these effects to the required level of
accuracy, we will include a dynamical strange and charm quarks into our
calculations. 

\phantom{TEST}

{\bf Acknowledgments:} We thank Gilberto Colangelo, Achim Denig, Fred
Jegerlehner, Harvey B. Meyer and Rainer Sommer for useful discussions. We are grateful to our colleagues within the CLS project for sharing gauge ensembles. Calculations of correlation functions were performed on the dedicated {Q}{C}{D} platform "Wilson" at the Institute for
Nuclear Physics, University of Mainz. This work was supported by DFG (SFB443) and
the Research Center EMG funded by Forschungsinitiative Rheinland-Pfalz.

\end{document}